# The Green Management Towards a Green Industrial Revolution

Malgorzata Rutkowska[1][0000-0002-0305-5555] and Adam Sulich[2][0000-0001-8841-9102]

[1] Wrocław University of Science and Technology, ul. Wybrzeze S. Wyspianskiego 27
50-370 Wroclaw, Poland
`malgorzata.rutkowska@pwr.edu.pl`
[2] Wroclaw University of Economics and Business; ul. Komandorska 118/120;
53-345 Wrocław, Poland
`adam.sulich@ue.wroc.pl`

**Abstract.** The Green Management (GM) is now one of many methods proposed to achieve new, more ecological, and sustainable economic models. The paper is focused on the impact of the developing human population on the environment measured by researched variables. Anthropopressure can have both a positive and a negative dimension. This paper aims to present an econometric model of the Green Industrial Revolution (GIR) as a result of GM. The GIR is similar to the Fourth Industrial Revolution (FIR) and takes place as the next stage in the development of humanity, especially in the perception of both machines and devices and the natural environment. The processes of the GIR in the European Union can be identified based on selected indicators of Sustainable Development (SD), in particular with the use of indicators of the Green Economy (GE) using taxonomic methods and regression analysis. The GM strives to implement the idea of the SD in many areas, to transform the whole economy and the elements of this process are visible as the technical progress. The adopted direction of economic development depends on the assumptions of strategic management, decisions, and their effect identify as GM.

## 1 Introduction

In the modern world, environmental pollution is one of the side effects of economic development. And so far in all countries worldwide, both economic growth and development are based mostly on the exploitation of natural resources. This is the case despite successive pro-ecological programmes and strategies developed both at the level of enterprises and entire countries and their associations [18]. The human population has a growing impact on the natural environment and this impact is called anthropopressure [3, 11, 15]. However, it was not until the 1980s that the concept of Sustainable Development (SD) was proposed. The SD was defined as "a method of management in which meeting the needs of the present generation will not reduce the chances of meeting the needs of future generations" [2]. Besides the negative anthropopressure, there is quite a big group of human activities towards the protection of the natural environment, which is a positive anthropogenic pressure [3]. This includes activities aimed at restoring or maintaining natural habitats, restoring and protecting species, forests and limiting the consumption of natural resources [12]. This type of combined actions and processes can be recognized as the Green Management (GM) and it is associated with the economic transformation process called the Green Industrial Revolution (GIR). The change of the paradigm of the role of human actions, machines, and devices in environment protection is visible. In the literature, there is a lot of taxonomic linear ordering methods however these methods are not further used or complemented by linear regression or statistical methods. Therefore, we identified two types of research gaps. First related to the method itself, second expressed in the question: *How to represent real economy changes induced both by the ecological concerns and technological progress*? The described research gap indicates developed in this paper discussion synthesis axis.

This paper aims to present the phenomenon of the GIR in the European Union countries and propose its econometric model. We understand that the GM is one of the ways to effectively implement pro-ecological solutions and to implement the so-called green practices [4] that will contribute to reducing the negative impact of the organization itself on the environment, its processes, and products perspectives. The most recent studies [10] on GM define its concept as a practice that results in environmentally friendly products and minimizes environmental impact through green production, green research and development, and green marketing [10]. These elements are based on the technological development and patents related to environmental protection [1, 13, 15].

This paper is organized as follows. After this introduction, as the first section of the paper, a literature review is presented. The second part is dedicated to the variables used in the next chapters of this paper. The third part consists of the description of used methods, followed by the results and their discussion. The conclusion describes the author's final remarks and managerial implications, along with the limitations and further research possibilities.

## 2   Industry 4.0 and Green Industrial Revolution

The analysis of the industry development leads to the observation that this process is based on the change of the vision or approach of the technology and environment. Figure 1 presents the development of the new technologies as successive stages. The first is Industry 1.0. It was the revolution that began in 18th century England and was the transition from a manufacturing economy to large-scale factory production. The crucial point was the invention and implementation of the steam engine brought production into the era of industrialization. The second stage is Industry 2.0 that started at the end of the 19th century and brought a rapid development of science and, as a result, new technical solutions. Electrification appeared and it was electricity that replaced the steam engines, and production lines could produce goods in large series. The next stage is Industry 3.0 that just originated after the Second World War, and with it, the scientific and technical revolution (also known as the third industrial revolution) began. The efficient computers and data processing systems made it possible to control machines using the software. As a result, the machines gained greater efficiency, precision, and flexibility, and the digitization process made it possible to achieve higher and higher degrees of automation. Planning and control systems began to emerge, the purpose of which was to coordinate activities within production.

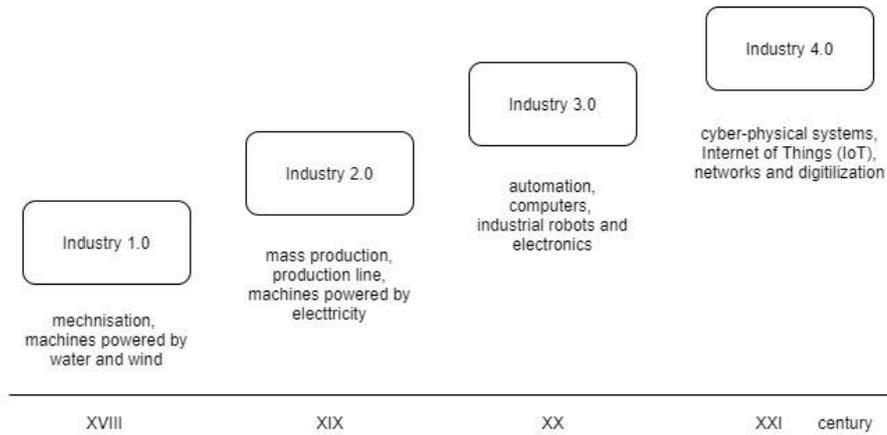

**Fig. 1.** Technological progress stages. Source: own elaboration based on [14].

The last stage is Industry 4.0 (the Fourth Industrial Revolution) which is a synonym for the development of high technologies. Systems integration and networking are taking place here. Industry 4.0 integrates people and digitally controlled machines with the Internet and information technologies. The materials produced or used for production are always identifiable and can communicate independently with each other. The information flow is vertical: from individual components to the company's IT department and from the IT department to the components. The second direction of information flow is carried out horizontally: between the machines involved in the production process and the production system of the enterprise. Therefore, Industry 4.0 should be understood as a collective term for innovative technologies operating in conjunction with the concept of value chain organization. As part of these processes, in modularly structured smart factories, cyber-physical systems monitor physical processes, create virtual copies of the physical world, and make decentralized decisions. "Through the Internet of Things, cyber-physical systems communicate with each other and cooperate and with people in real-time. Through the Internet of Services, both internal and inter-organizational services are offered and used by participants in the value chain" [8]. It is also important that Industry 4.0 is not only about technology but also about new ways of working and the role of people and their devices and the natural environment. The concept of Industry 4.0 means the unification of the real world of production machines with the virtual world of the Internet and information technology.

One of the most important Chinese economic strategists is H. Angang (2017), who believes that the history of mankind can be divided into three main stages, where:

1st. until the industrial revolution - man was a slave to the forces of nature subordinated to its enormous forces;
2nd. encompassing the golden age of industrialization, urbanization, and modernization - it was the other way around - it was the man who tried to impose his will and vision on nature, leading to what should be called the black industrial revolution;
3rd. which is inevitable, in which "humanity is no longer the master of nature, but its friend".

Such a division is related to the four successive stages of the industrial revolution and is similar to those presented in Figure 1. It is possible to consider what is the relationship between the green economy and Industry 4.0. It should be said that this is a conscious activity in which humanity tries to find a way out of ecological abnegation. It becomes important to mobilize all known techniques to completely rebuild the industrial society, to again change the perspective and view on the surrounding natural world (environment). Therefore, one should rather talk about the Green Industrial Revolution (GIR) and the so-called Green Industry 4.0. Therefore, it is helpful to apply a circular and Green Economy (GE) in which biological cycles are mimicked (Fig. 2). The idea of a closed circuit is based on the so-called incorporating sustainable thinking at every stage of working with a product or service, which thus indicates the longevity or even immortality of products.

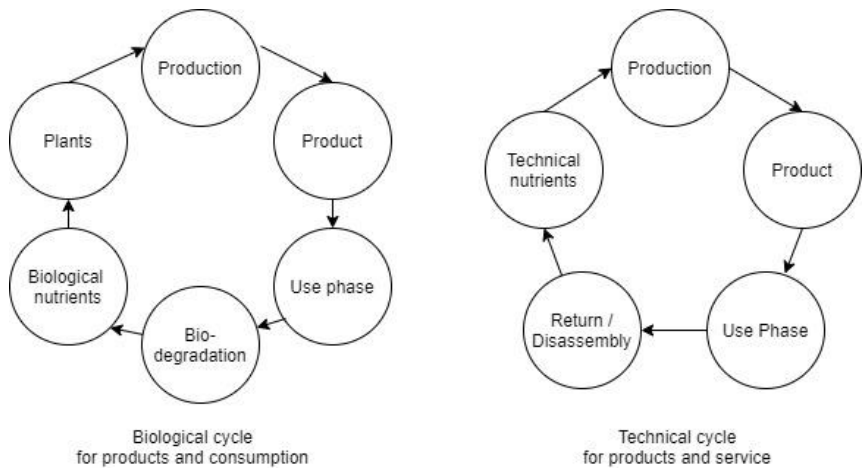

**Fig. 2.** Circular models of compared cycles. Source: own elaboration based on [5]

There are many similar elements in both cycles presented in Figure 2. A technical cycle consists of five elements when biological is based on the six consecutive elements. The main difference is the biodegradation process which is an additional step in the biological cycle. However, the GIR proposition is to add the recycling and reuse processes.

## 3   Taxonometric Analysis

The evaluation of the Green Industrial Revolution (GIR) among the methods of measuring the effects of green management or pro-ecological strategies of countries [9] and local government units or companies the reference method [7] or its modification [16, 17] is often used.

The reference method comes down to the determination of a synthetic variable being a function of the normalized features of the data input set. The method is also called the information capacity indicator method. The essence of this method lies in the procedure according to which, from the explanatory variables in the matrix, a combination of variables is selected. The so-called integral information capacity is greatest. Moreover, this method allows to measure and compare variables of different sizes and dimensions with each other, because data standardization procedure is used.

The purpose of the method is to compare the level of development described as the GIR at the international level and those creating a circular economy in the EU [6]. The indicators were defined by Eurostat and their units are presented below (Table 1). The variables used in calculations were assigned by symbols x with the number lower index ($x_i$). As result, the total number of 12 variables was determined in this way.

**Table 1.** Green Growth Indicators based on the Eurostat method

| Green Growth area | Measured characteristic | Symbol |
| --- | --- | --- |
| Production and consumption | Generation of municipal waste per capita [kg/person] | $x_1$ |
| | Generation of waste excluding major mineral wastes per GDP unit [kg/thousand euro] | $x_2$ |
| | Generation of waste excluding major mineral wastes per domestic material consumption [%] | $x_3$ |
| Waste management | The recycling rate of municipal waste [%] | $x_4$ |
| | The recycling rate of all waste excluding major mineral waste [%] | $x_5$ |
| | The recycling rate of e-waste [%] | $x_6$ |
| | Recycling of biowaste [kg/person] | $x_7$ |
| | The recovery rate of construction and demolition waste [%] | $x_8$ |
| Secondary raw materials | Circular material use rate [%] | $x_9$ |
| | Trade in recyclable raw materials [tones · $10^{-3}$] | $x_{10}$ |
| Competitiveness and innovation | Private investments, jobs, and gross value added related to circular economy sectors [bn. euro $10^{-3}$] | $x_{11}$ |
| | Patents related to recycling and secondary raw materials [number] | $x_{12}$ |

Source: Own study based on [6].

Secondary data from the year 2019 collected by Eurostat [6] were used for the calculations, which ensures the comparability and reliability of the data. The raw source data for Each EU-28 country are enclosed in the appendices section (table A). The rational reason for the choice of the taxonomic method, especially the zero unitarization method [7] is the development of an econometric model of the GIR and the indication of its determinants in the further part of this paper. Moreover, the application of the standard method allows for the verification of the obtained results in the comparison of countries with similar development described in the literature [9]. Since the set of independent features contains variables that cannot be aggregated directly using appropriate standardization, normalization formulas were applied. Among the formulas, the method of zero unitarization was selected to standardize the process based on the interval of a normalized variable. Variables that positively influence the described phenomenon are called stimulants. Their opposites are de-stimulants. Indicators are selected to a standardization process based on the following formulas:

$$\text{for stimulants: } z_{ij} = \frac{x_{ij} - \min(x_{ij})_i}{\max(x_{ij})_i - \min(x_{ij})_i} \quad (1)$$

$$\text{for de-stimulants: } z_{ij} = \frac{\max(x_{ij})_i - x_{ij}}{\max(x_{ij})_i - \min(x_{ij})_i} \quad (2)$$

where:
$z_{ij}$ – is the normalized value of the *j*-th variable in the *i*-th country;
$x_{ij}$ – is the initial value of the *j*-th variable in the *i*-th country.

Diagnostic features normalized in an above-mentioned way take the value from the interval [0;1]. The closer the value to unity, the better the situation in terms of the investigated feature, and the closer the value to zero, the worse the situation.

In the next step, the normalized values of variables formed the basis for calculating the median and standard deviation for each of the countries studied. Median values were determined using the formula:

$$\text{for even number of observations: } Me_i = \frac{Z\left(\frac{m}{2}\right)_i + Z\left(\frac{m}{2}+1\right)_i}{2} \quad (3)$$

$$\text{for odd number of observations: } Me_i = Z\left(\frac{m}{2}+1\right)_i \quad (4)$$

where:
$z_{i\,(j)}$ – is the *j*-th statistical ordinal for the vector $(Z_{i1}, Z_{i2}, \ldots, Z_{im})$, i = 1, 2, …, n; j = 1,2, …, m.

In turn, the standard deviation was calculated according to the following

$$S_{di} = \sqrt{\frac{1}{m}\sum_{j=1}^{m}(z_{ij} - \bar{z})} \quad (5)$$

Based on the median and standard deviation, an aggregate measure $w_i$ of the green management towards the GIR was calculated for each country:

$$w_i = M_{ei}(1 - S_{di}); w_i < 1 \quad (6)$$

Values of the measure closed to one indicate a higher level of the greening of industrial parts of the economy in the specific member state, resulting in a higher rank. The

47aggregate measure prefers countries with a higher median of features describing the specific country and those with smaller differentiation between the values of features in the specific country expressed as standard deviation [9]. The procedure chosen for evaluating the GIR provided a multidimensional comparative analysis. Such analysis allowed for a comparison between member states of the EU providing grounds for classifying them into uniform groups (Table 2).

**Table 2.** Green Growth Indicators aggregate measure comparative analysis

| Group | Mathematical characteristic | Meaning |
|---|---|---|
| I | $w_i \geq \bar{w} + S$ | high level |
| II | $\bar{w} + S > w_i \geq \bar{w}$ | medium-high level |
| III | $\bar{w} \geq w_i \geq \bar{w} - S$ | medium-low level |
| IV | $w_i < \bar{w} - S$ | low level |

Where $\bar{w}$ is the mean value of the synthetic measure; S is the standard deviation of the synthetic measure

According to the $w_i$ values the EU countries were assigned to one of the groups concerning their level of evaluating the GIR expressed in the greening economy.

**Table 3.** Groups of the EU countries with a similar level of Green Industrial Revolution

| Group | Countries |
|---|---|
| I | Finland, Denmark, Sweden, Germany, Austria, Estonia, Latvia, United Kingdom, |
| II | Luxembourg, The Netherlands, Lithuania, Belgium, France, Czechia, Slovakia, Slovenia, Ireland |
| III | Poland, Hungary, Malta, Cyprus, Italy, Spain, Portugal, Greece |
| IV | Romania, Bulgaria, Croatia |

Source: Authors' calculations

The level of the GIR was evaluated in the 28 EU countries based on 12 variables (Table 1) and the results of the analysis were presented in Table 3. The analysis shows that there are countries that operate in high-level management towards the GIR and it presents countries with the best conditions for sustainable (green) development described at the international level.

## 4 Model of the Green Industrial Revolution Proposition

The need to monitor the level of development of the GE and the FIR in the EU is an element of the assessment of the GIR. Therefore, an econometric model of the GIR was proposed based on the secondary data monitored by Eurostat in 2019 (Table 1). The variables used to propose the model were the same. The determinants of the model are the basic tool for both diagnosis and observation of the development of the GE and Industry 4.0 in various areas of socio-economic life. The model was developed using the regression method in the Statistica® programme. The dependent variable was the number of patents related to recycling and recyclable materials $x_{12}$ (Table 1). The reason for choosing the described variable is the impact of technology on environmental protection, which may be measured, inter alia, by the number of patents. The proposed mathematical model is a data description.

The study covered 28 European Union countries due to 12 indicators as their features (Table 1). Raw data presented in Table A in the appendix section. The methods of regression and correlation were used to create the regression model. Correlation is the study of the strength of the relationship between variables, and regression determines the shape of this relationship. Eurostat found that circular economy indicators also correspond to the concept of the GE [6]. Moreover, there is a cause-and-effect relationship between the indicators (variables). The study of the relationships between many independent variables ($x_1$-$x_{11}$) and one dependent variable ($x_{12}$) is called multiple regression.

Table 6 shows the results of the regression procedure for the variable number of patents related to recycling and secondary raw materials. The computer programme used the full number of countries (cases) under consideration N= 28. The basic measure of the theoretical and experimental curve fitting is the coefficient of determination $R^2 \approx 0.88$ (equal to the square of the correlation coefficient $R \approx 0.88$). The closer the absolute value | R | is to 1, the correlation is stronger.

**Table 4**. Dependent Regression Summary - Number of Patents

| N=28 | b* | Std. Err b* | b | Std. Err b | t(15) | p |
|---|---|---|---|---|---|---|
| Intercept | | | 9,85358 | 39,40932 | 0,25003 | 0,805955 |
| x1 | -0,02396 | 0,179326 | -0,00367 | 0,02745 | -0,13361 | 0,895489 |
| x2 | 0,83166 | 0,241834 | 0,13312 | 0,03871 | 3,43896 | 0,003654 |
| x3 | -1,15224 | 0,297816 | -3,34653 | 0,86497 | -3,86895 | 0,001514 |
| x4 | 0,33587 | 0,251141 | 0,36087 | 0,26983 | 1,33736 | 0,201032 |
| x5 | 0,21358 | 0,179599 | 0,19962 | 0,16786 | 1,18922 | 0,252838 |
| x6 | 0,02814 | 0,210937 | 0,06793 | 0,50921 | 0,13341 | 0,895644 |
| x7 | -0,22880 | 0,153723 | -0,32462 | 0,21810 | -1,48842 | 0,157362 |
| x8 | -0,30414 | 0,270217 | -0,13541 | 0,12031 | -1,12555 | 0,278043 |
| x9 | -0,07820 | 0,179469 | -0,12157 | 0,27901 | -0,43574 | 0,669233 |
| x10 | 0,82916 | 0,268304 | 2,86924 | 0,92845 | 3,09036 | **0,007463** |
| x11 | 0,44654 | 0,155048 | 0,00002 | 0,00001 | 2,88002 | 0,011449 |
| x12 | 0,43033 | 0,171987 | 0,00103 | 0,00041 | 2,50213 | 0,024403 |

Model details: R= 0,88137228; R^2= 0,87681709; Corrected $R^2$= 0,59827077;
F(12,15)=4,3508 p<0,00044 Std Err. Estim.: 0,185

The table above shows that the number of patents ($x_{12}$) is influenced by the variables marked in bold ($x_{10}$). The highest computer level of significance of *p* coefficients is shown by variable $x_{10}$, therefore this index was included in the linear equation (a value lower than 0.05 proves the significance of the coefficients and is marked by the Statistica® programme in bold). Corrected $R^2 \approx 0.5982$ is the coefficient of determination for small trials. F (12.15) ≈ 4.35 is the value of the F statistic, while *p* <0.00044 is the computer significance level for the F-test. The standard error of estimation ≈ 0.185 denoted by the symbol $u_i$ in the equation of the form:

$$y_i = a_i x + a_0 + u_i \qquad (7)$$

**Table 5.** Summary of the statistics obtained results

| Statistic feature | Value |
|---|---|
| R multiple | 0,881372279 |
| MultipleR2 | 0,776817095 |
| CorrectedR2 | 0,59827077 |
| F(12,15) | 4,35078738 |
| p | 0,00044596 |
| Error std. estimation | 0,18548170 |

The predicted values of the $x_{10}$ variable differ from the empirical values by an average of 0.185% (error standard estimation value). In equation 7, the coefficients $a_1$ and $a_0$ are listed in column b of Table 4, while column b* contains standardized regression coefficients. The relationship of the number of patents related to recycling and secondary raw materials ($x_{12}$) depends on $x_{10}$ - the share of recycled materials in the demand for raw materials. The regression equation has the form:

$$x_{12} = 0{,}553 \cdot x_{10} - 0{,}065 \pm 0{,}66349 \qquad (8)$$

This means that the growing share of recycled materials means that the number of patents related to recycling and secondary raw materials is increasing by 0.553(%). This model explains about 60% of the variability in the number of patents.

Then, the obtained results were verified using the Statistica® programme, in which a residual normality plot was prepared (Figure 3) is the result of the verification procedure for the obtained regression model (Table 6).

**Table 6.** Summary of test values verifying the obtained equation

| Characteristics | Durbin-Watson | Serial correlation |
|---|---|---|
| Values | 1,642365 | 0,176710 |

Source: Authors calculations

In Table 6 values of Durbin-Watson and serial correlation, statistics prove that there is no residuals autocorrelation.

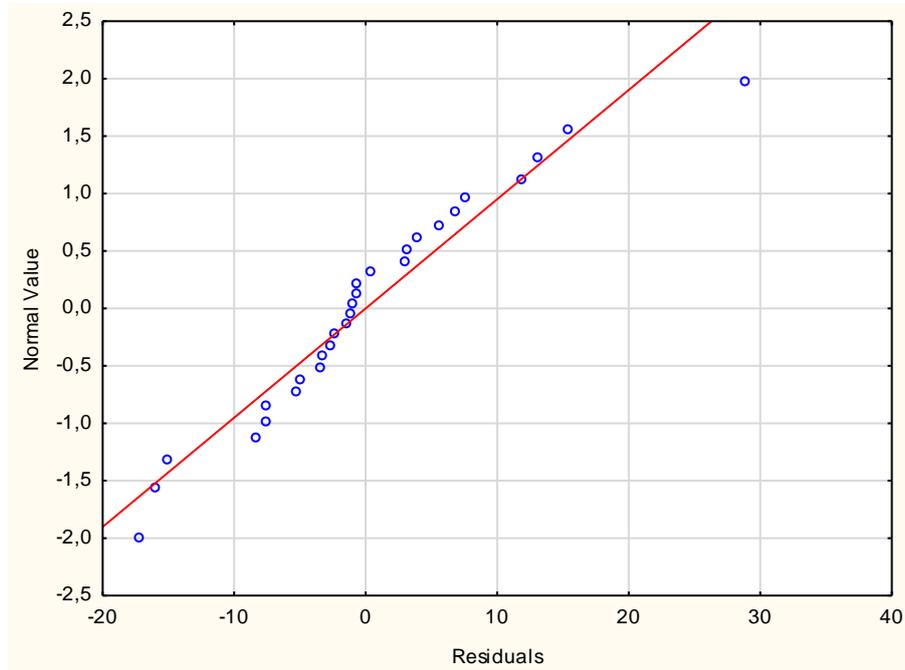

**Fig. 3**. Residual distribution normality plot. Source: own study results

The proposed econometric model of the GIR measured by the number of patents related to recycling and secondary raw materials assumes that the share of recycled materials in the demand for raw materials has the greatest impact on their quantity.

## 5   Conclusions

Based on the above discussion and results, the following conclusions were drawn. Firstly, linking the Fourth Industrial Revolution (Industry 4.0) with the Green Economy will contribute to a Sustainable Development and sustainable economy, and consequently an environmentally friendly economy. Secondly, the concept of Industry 4.0 is used in industry, research, and education and has an impact on the perception of the natural environment. More importantly, the changes brought about by the FIR transcend far beyond industry and affect the entire economy. The processes of transformation have a convergent character, use all mentioned concepts, and can be called the Green Industrial Revolution.

Industrial civilization, and hence the GIR, must be associated with the economy focused on SD with the GE. These links are particularly evident in the field of recycling and secondary raw materials patents. The models of biological and technical cycles represent this relationship.

The most exciting breakthroughs of the 21st century are taking place because of technological development, but also because of the expanding concept of a GE, and more precisely, the attempt to answer the question: how to combine environmental protection with economic development. The answer identified in this paper is related to the patents and the sector of environmental technologies development, which induces changes in the labour market in form of eco-innovations and green jobs.

On the other hand, the presented assumptions open new research avenues for green jobs and related to the eco-innovations, which turn into patents. There is a need to indicate the correct relationship between the labour market and the natural environment. Moreover, GM aims to both reduce the consumption of natural resources, generate less waste, reduce greenhouse gas emissions and, above all, eliminate social inequalities. The growing number of patents related to recycling and secondary raw materials and the growing popularity of ecological products and services are a manifestation of the growing influence of the GIR on business practice. Together with the development of environmental technologies, the changes towards the GIR can be observed and measured.

## Acknowledgments

The project is financed by the Ministry of Science and Higher Education in Poland under the program "Regional Initiative of Excellence" 2019–2022, project number 015/RID/2018/19, total funding amount 10,721,040.00 PLN.

The project is financed by the National Science Centre in Poland under the programme "Business Ecosystem of the Environmental Goods and Services Sector in Poland" implemented in 2020-2022 project number 2019/33/N/HS4/02957 total funding amount 120 900,00 PLN.

# Appendix

**Table A.** Indicators of the Green Industrial Revolution for individual European Union countries in 2019 - non-standardized values

| Country symbol | $x_1$ | $x_2$ | $x_3$ | $x_4$ | $x_5$ | $x_6$ | $x_7$ | $x_8$ | $x_9$ | $x_{10}$ | $x_{11}$ | $x_{12}$ |
|---|---|---|---|---|---|---|---|---|---|---|---|---|
| BE | 412 | 97 | 26,1 | 53,5 | 78 | 81,5 | 33,8 | 80 | 95 | 18,3 | 289,5 | 2,8 |
| BG | 419 | 418 | 13,3 | 29,4 | 27 | 64,1 | 96,5 | 43 | 90 | 3,1 | 88,1 | 0,5 |
| CZ | 316 | 74 | 7,8 | 29,7 | 60 | 74,3 | 37,9 | 13 | 92 | 6,9 | 3,3 | : |
| DK | 781 | 35 | 7,2 | 45,6 | 61 | 73,9 | 43 | 138 | 90 | 8,4 | 116,3 | 2,3 |
| DE | 632 | 55 | 12 | 66,7 | : | 69,3 | 33,9 | 114 | : | 11,4 | 1621,2 | 28,6 |
| EE | 359 | 657 | 35,1 | 28,3 | 10 | 59 | 33,3 | 13 | 7 | 11,4 | 4,1 | : |
| IE | : | 35 | 8 | : | 41 | 67,5 | 46,1 | : | 96 | 1,9 | 40,2 | : |
| GR | 488 | 78 | 11,5 | 15,8 | : | 60,3 | 32,7 | 12 | 88 | 2 | 222,0 | 0,6 |
| ES | 456 | 62 | 17,2 | 30 | 46 | 68,4 | 35,6 | 53 | 79 | 7,5 | 2319,9 | 11,0 |
| FR | 516 | 46 | 13,4 | 40,7 | 54 | 65,5 | 32,2 | 93 | 71 | 18,7 | 643,5 | 21,3 |
| HR | 393 | 75 | 8,2 | 18 | 52 | 60,1 | 58,3 | 7 | 76 | 4,3 | 70,7 | 0,6 |
| IT | 486 | 69 | 22,5 | 44,3 | 68 | 66,8 | 32,1 | 86 | 98 | 16,6 | 683,9 | 17,8 |
| CY | 638 | 38 | 5,4 | 17,9 | 31 | 59,8 | 27 | 30 | 57 | 2,4 | 0,2 | 0,1 |
| LV | 404 | 97 | 9,1 | 28,7 | : | 53,9 | 23,1 | 24 | 98 | 5,4 | 29,2 | 0,2 |
| LT | 448 | 102 | 7,8 | 33,1 | 68 | 59,8 | 45,9 | 46 | 97 | 4,1 | 35,2 | 0,4 |
| LU | 607 | 32 | 11,4 | 47,4 | 64 | 67,9 | 42,5 | 111 | 100 | 9,6 | 23,5 | : |
| HU | 377 | 98 | 9,2 | 32,2 | 43 | 50,1 | 50,7 | 23 | 99 | 5,8 | 41,3 | 0,9 |
| MT | 606 | 63 | 9,3 | 6,7 | 43 | 37,1 | 13,1 | 0 | 100 | 7 | 0,1 | : |
| NE | 523 | 64 | 25,6 | 51,8 | 72 | 71,7 | 39,3 | 143 | 100 | 25,8 | 1049,9 | 5,2 |
| AT | 560 | 52 | 10 | 56,9 | 66 | 67,1 | 40,7 | 175 | 88 | 10,9 | 224,8 | 3,5 |
| PL | 286 | 183 | 11,8 | 32,5 | 56 | 57,6 | 33,1 | 17 | 91 | 11,6 | 236,0 | 4,7 |
| PT | 460 | 67 | 7,7 | 29,8 | 52 | 57,1 | 42,7 | 72 | 97 | 2,1 | 653,7 | 1,4 |
| RO | 247 | 140 | 4,7 | 13,2 | 30 | 55,9 | 22,5 | 18 | 85 | 1,7 | 27,1 | 1,1 |
| SL | 449 | 79 | 11,4 | 54,1 | 80 | 67 | 47,7 | 34 | 98 | 8,4 | 110,3 | 0,5 |
| SK | 329 | 100 | 11,8 | 14,9 | 44 | 64,3 | 40,3 | 24 | 54 | 5 | 9,0 | 0,6 |
| FL | 500 | 74 | 8,2 | 40,6 | 37 | 60,9 | 43,2 | 62 | 87 | 6,5 | 17,4 | 2,0 |
| SE | 451 | 50 | 9,2 | 47,5 | 49 | 71,8 | 51,6 | 70 | 61 | 6,8 | 674,3 | 4,1 |
| UK | 483 | 57 | 21,3 | 43,3 | 58 | 60,6 | 36,6 | 78 | 96 | 16,2 | 83,1 | 31,0 |

Source: own study based on secondary data [6].
Symbol: means no data. Country symbols by ISO 3166-1 alfa-2.